\documentclass[prx,twocolumn,preprintnumbers,showkeys,superscriptaddress]{revtex4-1}

\usepackage{graphicx}
\usepackage{ifthen}
\usepackage{hyperref}	
\usepackage{color}
\usepackage{amssymb}
\usepackage{wrapfig}
\usepackage{stmaryrd}
\usepackage{epsfig}
\usepackage{amsmath}

\begin{document}

\columnseprule = 0 pt

\preprint{New. J. Phys. \textbf{18}, 093042 (2016); \href{http://dx.doi.org/10.1088/1367-2630/18/9/093042}{doi:10.1088/1367-2630/18/9/093042}}

\title{Dissipative Preparation of Antiferromagnetic Order in the Fermi-Hubbard Model}

\author{J. Kaczmarczyk}
\email{jan.kaczmarczyk@ist.ac.at}
\affiliation{Institute of Science and Technology Austria, Am Campus 1, 3400 Klosterneuburg, Austria}

\author{H. Weimer}
\affiliation{Institut f\"ur Theoretische Physik, Leibniz Universit\"at Hannover, Appelstr. 2, 30167 Hannover, Germany}

\author{M. Lemeshko}
\email{mikhail.lemeshko@ist.ac.at}
\affiliation{Institute of Science and Technology Austria, Am Campus 1, 3400 Klosterneuburg, Austria}

%\date{\today}

\newcommand{\tr}{\textcolor{red}}
\newcommand{\tg}{\textcolor{green}}
\newcommand{\tb}{\textcolor{blue}}
\newcommand{\tm}{\textcolor{magenta}}
\newcommand{\bk}{\mathbf{k}}
\newcommand{\bQ}{\mathbf{Q}}
\newcommand{\ua}{\uparrow}
\newcommand{\da}{\downarrow}
\newcommand{\dg}{^\dagger}
\newcommand{\eq}{\begin{equation}}
\newcommand{\eqx}{\end{equation}}
\newcommand{\eqn}{\begin{eqnarray}}
\newcommand{\eqnx}{\end{eqnarray}}
\newcommand{\half}{\frac{1}{2}}
\newcommand{\ran}{\rangle}
\newcommand{\lan}{\langle}
\newcommand{\s}{\sigma}
\newcommand{\veci}{\mathbf{i}}
\newcommand{\vecj}{\mathbf{j}}
\newcommand{\rlan}{\rangle\langle}
\newcommand{\trtxt}[2][]{\text{Tr}_{#1}\{#2\}}

\begin{abstract}
The Fermi-Hubbard model is one of the key models of condensed matter physics, which holds a potential for explaining the mystery of high-temperature superconductivity. Recent progress in ultracold atoms in optical lattices has paved the way to studying the model's  phase diagram using the tools of quantum simulation, which   emerged as a promising alternative to the numerical calculations plagued by the infamous sign problem. However, the temperatures achieved using elaborate laser cooling protocols so far  have been too high to show the appearance of antiferromagnetic and superconducting quantum phases directly. In this work, we demonstrate that using the machinery of dissipative quantum state engineering, one can observe the emergence of the antiferromagnetic order in the Fermi-Hubbard model with fermions in optical lattices. The core of the approach is to add incoherent laser scattering in such a way that  the antiferromagnetic state emerges as the dark state of the driven-dissipative dynamics. The proposed controlled dissipation  channels described in this work are straightforward to add to already existing experimental setups.
\end{abstract}


%67.85.-d - Ultracold gases, trapped gases
%75.10.Jm - Hubbard model - magnetic ordering (quantized spin model)
%71.10.Fd - Lattice fermion models,


\maketitle{}

\section{\label{sec:intro}Introduction}

Experimental progress with ultracold fermions in optical lattices~\cite{Lewenstein2007,RevModPhys.80.885} leads the way to achieving one of the key goals of quantum simulation~\cite{RevModPhys.86.153} -- mimicking realistic condensed matter systems. To date, the experiments covered a broad range of systems and interaction regimes, from probing the BEC-BCS crossover in lattices \cite{Chin2006}, to the observation of a fermionic Mott insulator~\cite{Schneider2008,Jordens2008}, to studying short range magnetism~\cite{GreifScience13} and multiflavor spin dynamics \cite{KrauserNatPhys12}, to realizing topological Haldane model~\cite{JotzuNature14} and artificial graphene sheets~\cite{UehlingerPRL13}. These discoveries pave the way to use ultracold atoms to reveal the properties of the repulsive Fermi-Hubbard model~\cite{Esslinger2010, LeHur2009}. The latter is of particular importance since it represents a playground to get insight into  the physics of high-temperature superconductivity and related phenomena observed in the cuprates~\cite{Keimer2015}.

In the case of one particle per site and large on-site interaction, $U$, the Fermi-Hubbard model exhibits the transition to the Mott-insulating state~\cite{Schneider2008,Jordens2008} around the temperature $T \sim U$. If the temperature is decreased further and reaches the so-called `N\'{e}el temperature,' $T_{\rm N} \sim 4t^2/U$, where $t$ gives the hopping rate between neighboring sites,  the transition to the antiferromagnetic (AF) phase is expected~\cite{Hart2015,Esslinger2015}.  Currently the  temperatures achievable in experiment are slightly above the N\'{e}el temperature where AF correlations can already be observed, for instance, $T/T_{\rm N} \approx 1.42$ has been reached in Ref.~\cite{Hart2015}.
Ultimately, in order to study the superconducting phase or other phenomena related to pairing in high-temperature superconductors, the temperature needs to be substantially lower. Therefore, due to the experimental limitations inherent to the standard laser cooling techniques, it is crucial to develop alternative approaches~\cite{Ho2009,Bernier2009,Heidrich-Meisner2009, Diehl2010, Williams2010, Medley2011, PhysRevA.84.031402, Colome-Tatche2011, Lemeshko2012, Yi2012, PhysRevLett.111.033607, PhysRevLett.115.215301, 2015arXiv151004883M} to preparation of quantum phases in optical lattices.

In this work, we propose an efficient scheme for the preparation of antiferromagnetic order in the Fermi-Hubbard model, based on the ideas of dissipative state engineering, which have recently emerged in the context of many-particle systems~\cite{Diehl2008, PhysRevA.78.042307, PhysRevLett.101.200402, Verstraete2009, Weimer2010, Diehl2010, Diehl2011, Yi2012, PhysRevA.82.054103, PhysRevA.85.023604, Kordas2012, PhysRevLett.110.110502, PhysRevLett.111.033606, PhysRevLett.111.033607, Lemeshko2013, Weimer2013, PhysRevA.84.031402, Otterbach2014, PhysRevA.91.040302, PhysRevA.91.042117, Weimer2015b}
and have been implemented experimentally~\cite{Barreiro2010, PhysRevLett.95.073003, Barreiro2011, PhysRevLett.107.080503, Schindler2013b, Lin2013, Shankar2013}.
In such scenarios, a many-body state of interest (here: states exhibiting AF order) is prepared as a steady state of the quantum master equation governing the open system dynamics, as opposed to the ground state of the Hamiltonian. Such steady state can undergo quantum phase transitions to an ordered state of matter, which can be classified in close analogy to equilibrium systems~\cite{PhysRevLett.105.015702, PhysRevA.83.013611, PhysRevA.84.031402, PhysRevLett.110.233601, PhysRevLett.110.163605, PhysRevLett.110.195301, PhysRevLett.110.257204, Weimer2015b, Chan2015}.

We start with a system of fermions in an optical lattice as described by the Fermi-Hubbard model. The parameters of the Hamiltonian are left intact, instead we introduce dissipative channels on top of the unitary evolution.
As a result, fermions remain mobile in the optical lattice during the entire dissipative preparation stage, which should help with retaining coherence after the dissipation channels are switched off.
Furthermore, the dissipation channels of our scheme are implemented using the level structure of fermionic $^{40}$K, currently used in several laboratories~\cite{KrauserNatPhys12, GreifScience13, Esslinger2015, CheukPRL15, HallerNatPhys15, 2015arXiv151004744E}. Consequently, the presented scheme can be readily implemented into already existing experimental setups.

Theoretical description of open many-body quantum systems represents a challenging task and is currently an active field of research~\cite{PhysRevLett.115.080604,Weimer2015, Weimer2015b, PhysRevA.93.012106, PhysRevB.90.205125, Prosen2008, Dzhioev2015}. In our analysis of the dissipative Fermi-Hubbard model we use two complementary techniques: the Monte Carlo wave function (MCWF)~\cite{PhysRevLett.68.580,Molmer:93,PhysRevLett.70.2273} and the variational method~\cite{Weimer2015,Weimer2015b}, which is generalized here to the description of fermionic systems at half-filling. By using these two methods we demonstrate that a substantial AF magnetization is present in the system both for an exact solution on a  $3 \times 3$ lattice, as well as in the thermodynamic limit.

\section{\label{sec:model} The dissipative Fermi-Hubbard model}

We start with the Fermi-Hubbard Hamiltonian
\eq
\hat{H} = \sum_{\veci, \vecj, \s} t_{\veci \vecj} \hat{c}^\dagger_{\veci \s} \hat{c}_{\vecj \s} + U \sum_\veci \hat{n}_{\veci \ua} \hat{n}_{\veci \da},
\eqx
which has been experimentally realized in a range of systems such as $^6$Li~\cite{Hart2015} and $^{40}$K~\cite{Jordens2008}.
Our goal is to design dissipative processes in such a way that the state with an AF order is the dark state of the dissipative dynamics and the time evolution of the open system will drive it towards such a dark state.

The dynamics of an open quantum system is governed by the master equation for the system's density matrix
\eq
\dot{\rho} = -i \left[\hat{H}, \rho\right] +
\sum_{\veci, \vecj, \s, \alpha}{}^{'}
\left( \hat{j}^{(\alpha)}_{\veci\vecj, \s} \,\; \rho \,\; \hat{j}^{(\alpha)\dagger}_{\veci\vecj, \s} - \frac{1}{2} \left\{ \hat{j}^{(\alpha) \dagger}_{\veci\vecj, \s} \,\; \hat{j}^{(\alpha)}_{\veci\vecj, \s}, \rho \right\} \right),
\label{eq:master}
\eqx
where we set $\hbar \equiv 1$ and the primed sum runs over nearest-neighbor sites.
Since we start with a disordered sample, all possible nearest-neighbor configurations, including $| \ua; \ua \ran$, $| \da; \da \ran$, and $| \da\ua; 0 \ran$ will be present. The jump operators, therefore, need to convert the latter into those with the local antiferromagnetic order, $| \ua; \da \ran$.

\begin{figure*}[t!]
\begin{center}
\includegraphics[width=2.0\columnwidth,angle=0]{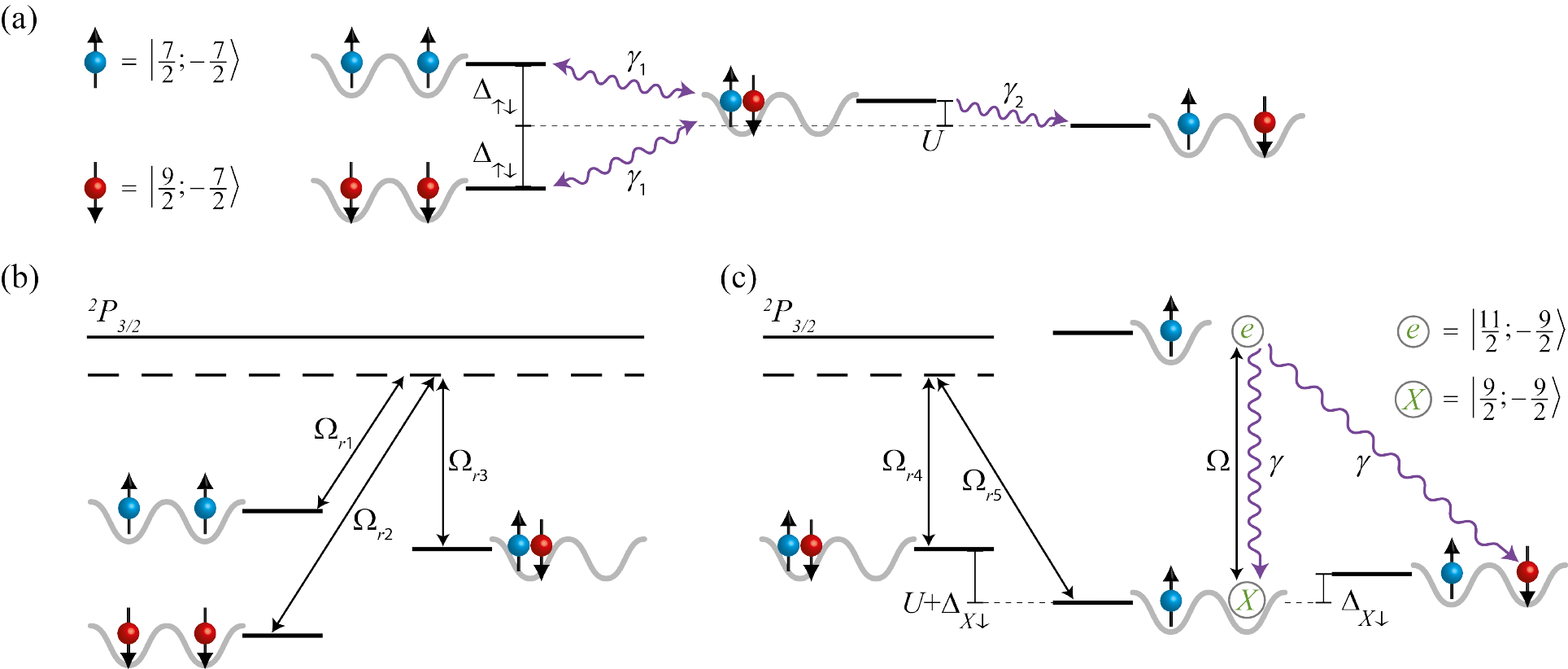}
\end{center}
\caption{Illustration of the dissipative processes. (a) Action of the jump operators on the nearest-neighbor sites; and (b)-(c) their implementation using Raman-assisted hopping. The Raman beams are labelled as $\Omega_{r1},...,\Omega_{r5}$,  the pumping beam by $\Omega$, and the decay rate is given by  $\gamma$; $\Delta_{\ua \da}$ gives the hyperfine splitting between the $| \ua \rangle$ and $| \da \rangle$ states, $\Delta_{X\da}$ gives the Zeeman splitting between the $| X \rangle$ and $| \da \rangle$ states, while $U$ denotes the on-site interaction between the $| \ua \rangle$ and $| \da \rangle$ states.
}
\label{fig1}
\end{figure*}

We choose the jump operators to be as follows:
\eqn
\hat{j}^{(1)}_{\veci\vecj, \ua} &=& \sqrt{\gamma_1} \, \hat{n}_{\veci \ua} (1-\hat{n}_{\vecj \da}) \hat{c}^\dagger_{\veci \da} \hat{c}_{\vecj\ua} , \qquad \hat{j}^{(2)}_{\veci\vecj, \ua} = \left( \hat{j}^{(1)}_{\veci\vecj, \ua} \right)\dg, \label{eq:jump1} \\
\hat{j}^{(1)}_{\veci\vecj, \da} &=& \sqrt{\gamma_1} \, \hat{n}_{\veci \da} (1-\hat{n}_{\vecj \ua}) \hat{c}^\dagger_{\veci \ua} \hat{c}_{\vecj\da} , \qquad \hat{j}^{(2)}_{\veci\vecj, \da} = \left( \hat{j}^{(1)}_{\veci\vecj, \da} \right)\dg, \\
\hat{j}^{(3)}_{\veci\vecj} &=& \sqrt{\gamma_2} \, (1-\hat{n}_{\veci \ua})\hat{n}_{\vecj \ua} \hat{c}^\dagger_{\veci \da} \hat{c}_{\vecj\da}, \label{eq:jump3}
\eqnx
where $\gamma_1$, $\gamma_2$ are the dissipation rates. The resulting dissipative dynamics is visualized in figure~\ref{fig1}(a): the jump operators $\hat{j}^{(1)}_{\veci\vecj, \sigma}$ (with $\s = \ua, \da$) turn the configurations on neighboring sites from $| \s; \s \ran$ to $|\ua\da; 0\ran$, whereas $\hat{j}^{(2)}_{\veci\vecj, \sigma}$ act in the opposite direction. These processes are labelled by the amplitude $\gamma_1$ in the figure.
Finally, the jump operators $\hat{j}^{(3)}_{\veci\vecj}$ turn the configurations $|\ua\da; 0\ran$, into those with the AF order, ${| \ua; \da \ran}$ (as  labelled by $\gamma_2$). As a result, the dissipative dynamics drives the system towards the AF phase.

We note that the jump operators $\hat{j}^{(3)}_{\veci\vecj}$ break the SU(2) symmetry, as the down-spin atom becomes more mobile. This, however, does not constitute a limitation of our scheme. Moreover, it is possible to reestablish the symmetry by using additional auxiliary states to induce hopping of the $|\ua\ran$-state atom away from the double-occupancy configuration.

As we discuss in the following section, such choice of the jump operators is straighforward to realize in experiment using incoherent laser scattering.

\section{Experimental implementation of the jump operators}

Simulating the Fermi-Hubbard model requires mapping the two spin states, $| \ua \ran$ and $| \da \ran$,  onto the fine or hyperfine components of the ground electronic state manifold of an ultracold atom. The on-site interaction between the spin components, $U$, can be tuned  using a Feshbach resonance~\cite{RevModPhys.82.1225}. We exemplify the scheme using the atomic level structure of fermionic ${}^{40}$K~\cite{Tiecke}, with the levels $| \ua \rangle \equiv |F=\frac{7}{2}; m_F=-\frac{7}{2}\rangle$ and $| \da \rangle \equiv |\frac{9}{2}; -\frac{7}{2}\rangle$. Furthermore, in order to realize the dissipative part of the dynamics, we introduce an auxiliary state,  $| X \rangle \equiv |\frac{9}{2}; -\frac{9}{2}\rangle$, belonging to the ${}^2 S_{1/2}$ manifold, as well as an electronically excited ${}^2 P_{3/2}$ state, $| e \rangle \equiv |\frac{11}{2}; -\frac{9}{2}\rangle$.

The first-stage jump operators, $\hat{j}^{(1)}_{\veci\vecj, \s}$ and $\hat{j}^{(2)}_{\veci\vecj, \s}$, can be implemented using Raman-assisted hopping, as illustrated in figure~\ref{fig1}(b). In such a process, transitions between two quantum states are induced using two laser beams, which are detuned from some excited state (here the ${}^2 P_{3/2}$ state). For example, the Raman beams $\Omega_{r1}$ and $\Omega_{r3}$ realize the Raman-assisted hopping between the $|\ua; \ua \ran$ and $|0; \ua \da \ran$ states. If the Raman beams, $\Omega_{r1,...,r3}$, are not phase-locked such hopping processes are dissipative.
Since the Raman-assisted hopping takes place directly between the initial and final states of the jump operators, the related dissipative processes are bidirectional. Therefore, we need to avoid populating the $|\ua; \da\ran$ state at this stage. Otherwise, the jump operators would also lead from the $|\ua; \da\ran$ state back to the $| \ua; \ua \ran$ and $| \da; \da \ran$ states. The on-site interaction energy $U$ (which we assume to be on the order of a few kHz) can be used for this purpose. We note that this step can be implemented in a coherent way as well, however, the required phase-locking of the lasers would introduce an additional complication into the experimental setup.

In order to implement the second-stage jump operators, $\hat{j}^{(3)}_{\veci\vecj}$, in a one-directional fashion, we use Raman-assisted hopping (lasers $\Omega_{r4}$ and $\Omega_{r5}$) from the $|\ua\da; 0\ran$ state to the $|\ua; X\ran$ configuration with an auxiliary $|X\ran$ state. This $|X\ran$ state is then pumped (laser $\Omega$) to an excited $|e\ran$ state, which can decay to the $|\da\ran$ state completing the process, as illustrated in figure~\ref{fig1}(c). The $|e\ran$ state cannot decay to the $|\ua\ran$ state because of selection rules on the $F$ quantum number. The Zeeman splitting, $\Delta_{X\da}$, and the energy, $U+\Delta_{X\da}$, differentiate  among the three states from the lower band in figure~\ref{fig1}(c). In order to resolve between these three states it is sufficient to use selection rules. To resolve between the $|\ua; X \ran$ and $|\ua X; 0 \ran$ states as the final states of the Raman-assisted hopping process we need nonzero on-site interaction between the $|\ua\ran$ and $|X\ran$ states. The typical values of the background scattering lengths, $a = 105 \, a_0$, in units of the Bohr radius $a_0$~\cite{PhysRevA.59.3660, Tiecke}, should be sufficient for this purpose. Spontaneous emission from the $| e \ran$ state ensures that the resulting dissipative processes are unidirectional and take place from the $|\ua \da; 0 \ran$ to the $|\ua; \da \ran$ state with AF ordering.

In total our dissipative preparation scheme requires six lasers ($\Omega_{r1}$, ..., $\Omega_{r5}$, and $\Omega$) in order to induce the dissipative transitions, in addition to the lasers  used to trap the atoms and prepare them in the $|\ua\ran$ and $|\da\ran$ states. Due to the $s$-wave character of the $^{2}S_{1/2}$ manifold, the optical lattice potential for the $|\ua \ran$, $|\da \ran$, and $|X\ran$ states will be the same, provided the optical-trapping lasers are sufficiently far-detuned from the $^{2}P_{3/2}$ states (cf. the discussion in Ref.~\cite{PhysRevA.70.033603}, section II.B).

The values of $\gamma_1$ and $\gamma_2$ are unrelated to the nearest-neighbor hopping integral $t$, however, all three of them are proportional to certain integrals involving two Wannier functions on the nearest-neighboring sites. As a safe estimate we consider $t$, $\gamma_1$, and $\gamma_2$ to be on the same order of magnitude.

Please note that the presented dissipative-preparation scheme is general and other states can be used as $|\ua \ran$, $|\da \ran$, $|X\ran$, and $|e\ran$. In particular, it should be possible to use the states of $^{40}$K, for which the Feschbach resonance is already known, i.e. $|\da \ran = |\frac{9}{2}; -\frac{9}{2}\rangle$, $|\ua \ran = |\frac{9}{2}; -\frac{7}{2}\rangle$ or $|\da \ran = |\frac{9}{2}; -\frac{9}{2}\rangle$, $|\ua \ran = |\frac{9}{2}; -\frac{5}{2}\rangle$. In the first case one could use $|X\ran = |\frac{7}{2}, -\frac{7}{2} \ran$ and $|e\ran = |\frac{11}{2}, -\frac{11}{2}\ran$, which requires a two-photon process from $|X\ran$ to $|e\ran$. This could be realized with Raman beams detuned from the $^2 D_{5/2}$ state or directly as a two-photon excitation. In the second case one could use $|X\ran = |\frac{9}{2}, -\frac{7}{2} \ran$ and $|e\ran = |\frac{9}{2}, -\frac{9}{2}\ran$ ($|e\ran$ is from the $^2P_{3/2}$ manifold), which would require resolving the $|X\ran \leftrightarrow |e\ran$ transition from $|\ua\ran \leftrightarrow  |\frac{9}{2}, -\frac{7}{2}\ran$.

\section{Time evolution and steady-state properties}

In order to reveal the properties of the system we use two complementary techniques: the Monte Carlo wave function (MCWF) technique~\cite{PhysRevLett.68.580,Molmer:93,PhysRevLett.70.2273}
on a $3\times 3$ lattice
and the variational method~\cite{Weimer2015,Weimer2015b} in the thermodynamic limit. While the latter was originally formulated for bosons, here  we extend it to fermionic systems.
In both methods we start from the Jordan-Wigner transformation in two spatial dimensions~\cite{JWtrafo,Fradkin1989,PhysRevB.43.3786,PhysRevB.48.6136}. The related Jordan-Wigner strings restrict the applicability of our variational scheme to the situation with one particle per site (half-filling). Experimentally, this is the most interesting regime as it corresponds to the maximal N\'{e}el temperature.

\subsection{Monte Carlo wave function}

We study the dynamics of the driven-dissipative system governed by equation~(\ref{eq:master}) using the MCWF method implemented in the QuTiP numerical library~\cite{qutip1,qutip2}.
We consider only the nearest-neighbor hopping $t_{\veci\vecj} \equiv -t$ (we use $t$ as the unit of energy hereafter) and study the time evolution of the half-filled $3 \times 3$ lattice as a function of the parameters $\gamma_1$, $\gamma_2$, and $U$. Since the lattice dimensions are odd numbers, we use the antiperiodic boundary conditions. This is required in the presence of the AF ordering because a particle hopping to its nearest neighbor across a boundary does not change the sublattice index, whereas for the same process within the boundaries such a change occurs. Therefore, when the hopping process takes place across the boundary, we introduce an additional spin-flip ($\hat{c}^\dagger_{\veci \s} \hat{c}_{\vecj \overline{\s}}$), which ensures that hopping processes within and across the boundaries are equivalent. We also introduce analogous corrections to the jump operators in equations~(\ref{eq:jump1})-(\ref{eq:jump3}).

The initial states for the MCWF realizations were chosen with randomly-positioned spin-up or spin-down particles (also allowing for double occupancies), however, the steady-state properties were found to be independent on the initial conditions.

\begin{figure}
\begin{center}
\includegraphics[width=\columnwidth,clip]{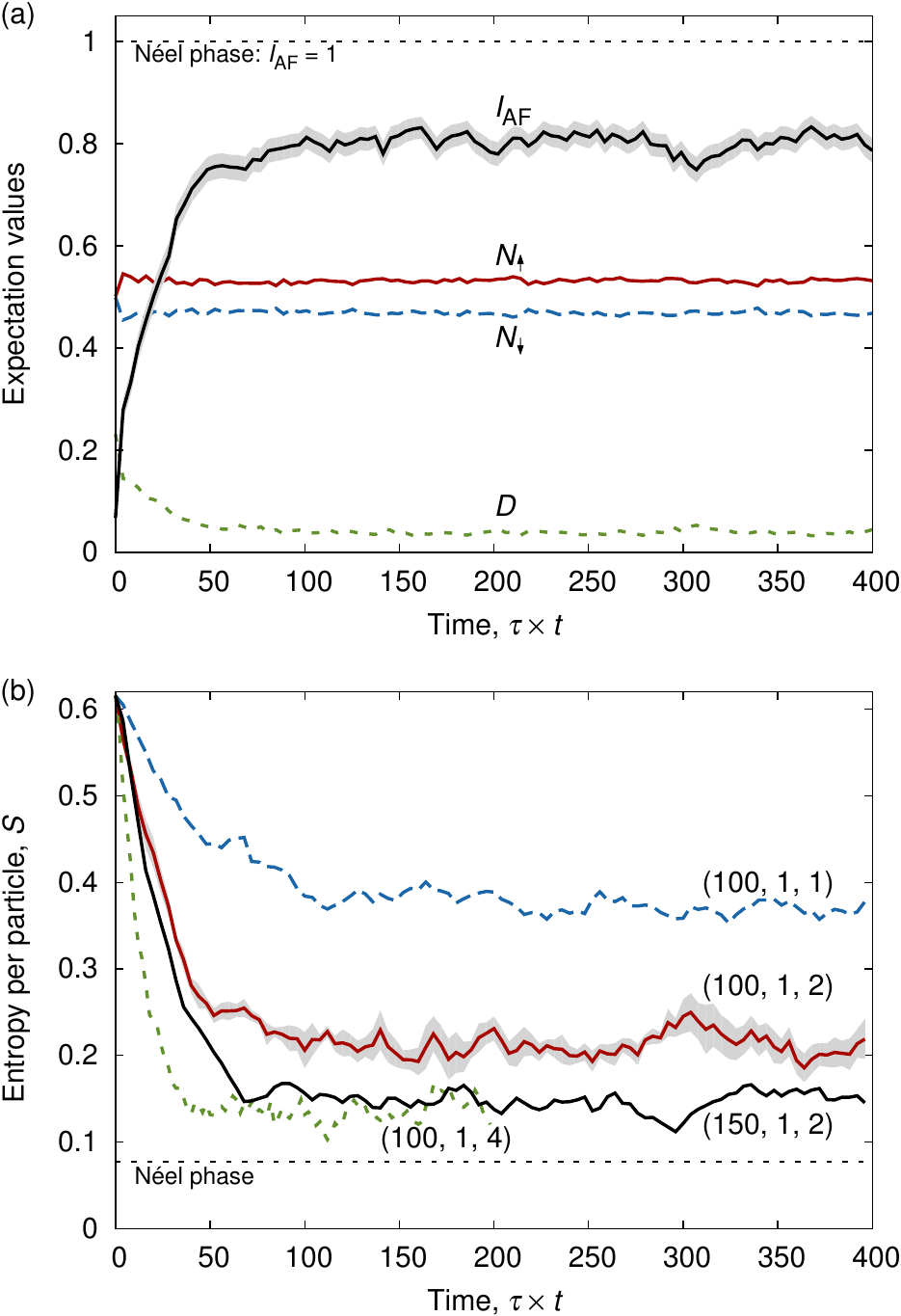}
\end{center}
\caption{Formation of the steady state. (a) Time-evolution of the total number of spin-up, $N_\ua$, and  spin-down, $N_\da$, atoms per site, the double occupancy probability, $D$, and the spin-structure factor, $I_{\rm AF}$, for $U = 100$, $\gamma_1 = 1$, and $\gamma_2 = 2$. (b) The von Neumann entropy per particle, $S$, for selected values of $(U, \gamma_1, \gamma_2)$, as labelled in the graph. The results have been averaged from 256 Monte Carlo realizations. The grey areas around selected curves correspond to one-sigma confidence interval of the results. The results ($I_{\rm AF}$ and $S$) for the N\'eel phase are represented by flat dashed lines.}
\label{fig2}
\end{figure}

The time evolution of the system's properties is shown in figure~\ref{fig2}. One can see that the steady state is reached for $\tau \times t \approx 50-100$, which corresponds to $\tau = 1-2$ seconds for $t = 50$ Hz.  Even for a large value of the  on-site interaction, $U = 100$, a small number of  double occupancies  is still present in the system, $D \approx 0.04$, see figure~\ref{fig2}(a). These states are involved in the dissipation processes as an intermediate step towards preparation of the AF ordered phase, cf. figure~\ref{fig1}(a). Non-zero double occupancies can lead to inelastic losses of atoms~\citep{Esslinger2010}, which however are more problematic for the attractive~\cite{PhysRevLett.92.040403}, than for the repulsive~\cite{Jordens2008} potassium gas. In the latter case  the inelastic decay time for atoms on doubly occupied sites was reported~\cite{Jordens2008} to exceed $850$~ms.  Consequently, such inelastic losses should not constitute a limitation of our scheme.

\begin{figure}
\includegraphics[width=\columnwidth]{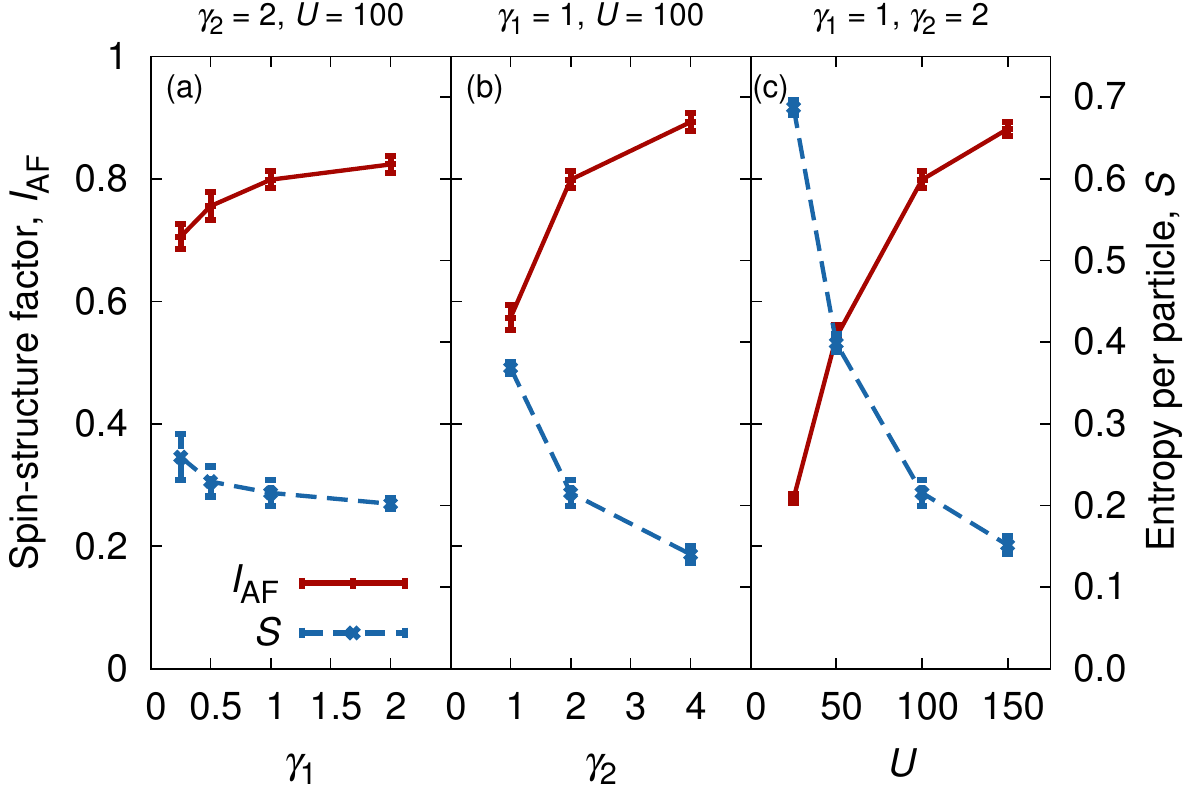}
\caption{ Steady-state properties: spin-structure factor, $I_{\rm AF}$, and entropy per particle, $S$, as a function of magnitudes of jump operators, (a) $\gamma_1$ and (b) $\gamma_2$, and (c) the on-site repulsion energy, $U$. The error bars of $I_{\rm AF}$ are calculated as the standard error for the results of single Monte-Carlo realizations, whereas those of $S$ are obtained using the standard deviation of the Monte-Carlo results averaged over time.}
\label{fig3}
\end{figure}

To quantify the AF ordering of the system, we evaluate the spin-structure factor, as defined by
\eq
I(\bQ) \equiv \frac{4}{N^2} \sum_{\mathbf{i}, \mathbf{j}} e^{i(\mathbf{R_i} - \mathbf{R_j})\bQ} \lan \s^{z}_{\mathbf{i}} \s^{z}_{\mathbf{j}} \rangle.
\eqx
Here, $\bQ$ is the ordering vector, which for the case of the AF phase is equal to $\bQ_{\rm AF} = [\pi, \pi]$, $\mathbf{R_i}$ and $\mathbf{R_j}$ are the lattice site vectors, $N$ is the number of particles ($N=9$ in our system), and $\sigma_{\veci(\vecj)}^z = \pm \frac{1}{2}$ is the $z$ component of the particles' spin.
In the case of the AF ordering, the relevant spin-structure factor is given by $I_{\rm AF} \equiv I(\bQ_{\rm AF})$.  For the steady-state it can be as large as $I_{\rm AF} \approx 0.8$ (cf.~also figure~\ref{fig3}), quite close to the fully polarized N\'eel state, for which $I_{\rm AF} = 1$. Note that the N\'eel state is an `ideal' AF state, to which the dissipative processes would drive the system in the limit of $\gamma_1, \gamma_2, U \gg t$. This is because the jump operators suppress intersite coherence in the system. However, the fact that fermions remain mobile in the optical lattice during the entire dissipative preparation stage should help with retaining coherence after the dissipation channels are switched off.

Figure~\ref{fig2}(b) illustrates the decrease of the system's entropy per particle, $S \equiv S_\text{tot} / N = - 1/N \, {\rm Tr} \rho \log \rho$, with time. Due to the large size of the system's density matrix $\rho$, in our numerical calculations we use the equivalent formula, $S_\text{tot} = - {\rm Tr} A \log A$. Here $A_{ij} = \lan \psi_i | \psi_j \ran$ is the matrix of overlaps of wave functions obtained from single realizations of the Monte Carlo algorithm. The resulting steady-state entropy per particle can be as low as $S \approx 0.1-0.3$ (cf. also figure~\ref{fig3}). For the N\'eel phase the entropy per particle is equal to $\frac{1}{9} \log 2~\approx~0.077$ due to the two-fold degeneracy corresponding to flipping of all spins. The relaxation time (usually below one second) increases with the increasing on-site interaction, $U$. For larger systems the relaxation times can be longer, due to possible formation of domains, as it is the case for coherent preparation strategies.

A slightly different number of spin-up, $N_\uparrow$, and spin-down, $N_\downarrow$,  atoms in the steady state is related to breaking of the $SU(2)$ symmetry by the jump operators $\hat{j}^{(3)}_{\veci\vecj}$. As a result, the spin-down atom becomes more `mobile'. Additionally, in the $3 \times 3$ lattice the number of sites is odd. Therefore, inherently in the steady state there is a spin-direction imbalance with $N_\ua > N_\da$. For a larger system, as well as for a system with even number of sites we would have $N_\ua \approx N_\da$. This, however, does not preclude the formation of the AF order.

Figure~\ref{fig3} shows the steady-state properties: the entropy per particle and spin-structure factor as a function of the parameters. In the employed range of parameters these features turn out to be strongly dependent on $\gamma_2$ and $U$, cf. figures~\ref{fig3}(b), (c), however, only weakly dependent on $\gamma_1$, cf. figure~\ref{fig3}(a). Furthermore, while $I_{\rm AF}$ grows substantially with increasing $U$ and $\gamma_2$,  for $\gamma_1$ a saturation effect is observed and increasing the magnitude above $\gamma_1 \approx 0.5$ does not improve the efficiency of the scheme significantly. These observations can be qualitatively understood from figure~\ref{fig1}(a). Namely, when the system is close to the AF phase most of the nearest-neighbor configurations are of the $|\ua; \da \ran$ type. The processes that drive the system away from the ordered state are related to coherent hopping from the $|\ua; \da \ran$ state to the  $|\ua \da; 0 \ran$ state. In the  large-$U$ limit, the timescale of such processes is given by $4 t^2/U$. Therefore, increasing $U$ reduces the contribution of the processes that destroy AF ordering. Increase of $I_{\rm AF}$ with $\gamma_2$ is expected, as the related dissipative processes drive the system directly into the AF ordered state. The saturation effect for $\gamma_1$ can be due to the bidirectional character of the related dissipative processes. For sufficiently large $\gamma_1$, the value of the spin-structure factor is determined by an interplay between the dissipative processes related to $\gamma_2$ and coherent hopping processes with a time scale governed by $4 t^2 / U$.

While the values of $U$ required for an efficient preparation of the AF order are quite large, they are within experimental reach, e.g. $U/t = 180$ was reported in Ref.~\cite{Jordens2008}. Increasing $U$ even further might lead to appearance of non-standard terms on top of the Fermi-Hubbard model~\cite{Dutta2015}.

\subsection{Variational scheme}

In order to describe the steady-state properties in the thermodynamic limit, we use a recently introduced variational principle~\cite{Weimer2015,Weimer2015b}. In this method, one has to minimize a suitable variational norm \footnote{cf. Refs.~\cite{Weimer2015,Weimer2015b} for details including the discussion of the choice of the variational norm.} of the master equation (\ref{eq:master})
\eq
||\dot{\rho}|| = ||-i[H,\rho] + \mathcal{D}(\rho)||.
\label{eq:min}
\eqx
Here $\mathcal{D}$ is the dissipative part as given by equation~(\ref{eq:master}).
The variational method was originally formulated for bosonic systems where a local ansatz on the density matrix can be used. For fermionic systems such a procedure is not possible directly. However, the Jordan-Wigner transformation~\cite{JWtrafo,Fradkin1989,PhysRevB.43.3786,PhysRevB.48.6136} can be used to map the fermionic creation and annihilation operators to spin operators (see appendix \ref{app:A} for details). The resulting system of spin-$1/2$ particles is equivalent to a system of hard-core bosons, however, with the fermionic statistics included. Namely, the equivalence with the starting fermionic system is ensured by long-range interaction terms -- the so-called Jordan-Wigner strings -- manifesting the anticommutation relations of the original fermionic particles. To apply the variational method we first need to approximate the `non-local' parts of the Jordan-Wigner strings and, thereby, transform the master equation~(\ref{eq:master}) to the form with at most two-site interactions (see appendix \ref{app:B} for details of this procedure).

\begin{figure}
\begin{center}
\includegraphics[width=\columnwidth]{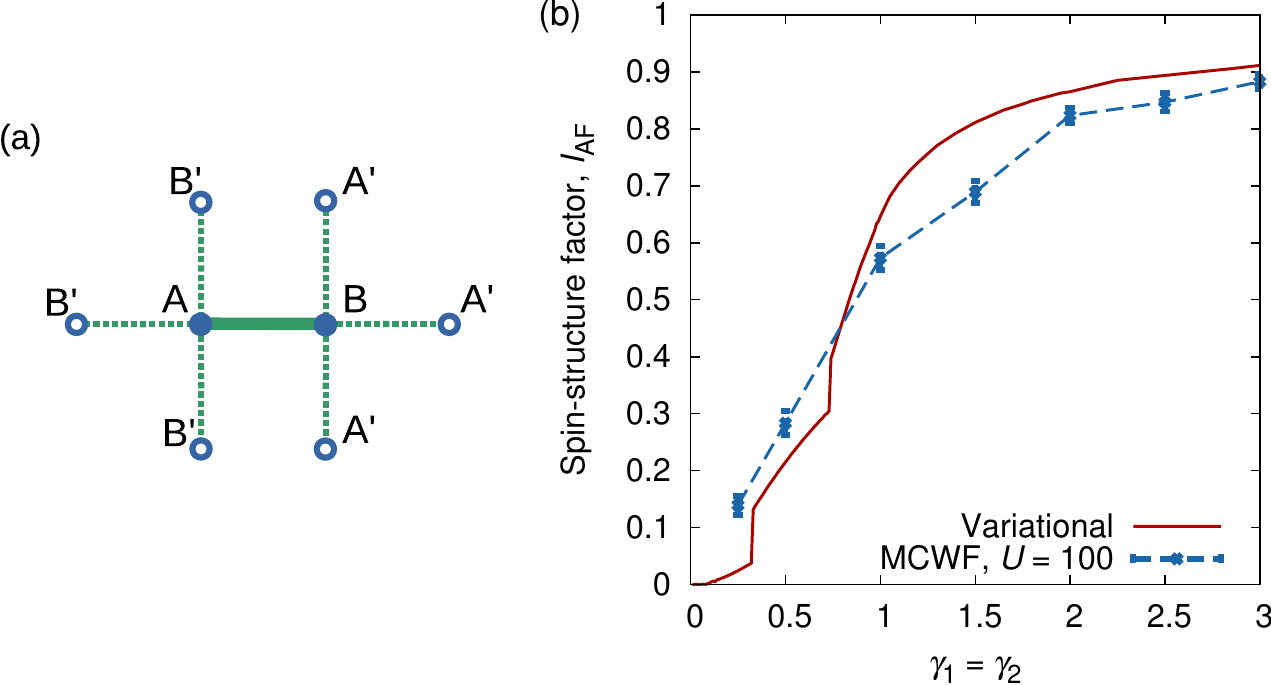}
\end{center}
\caption{Results in the thermodynamic limit. (a) Single link and the surrounding sites in the presence of  AF order. (b) The steady-state magnitude of the AF spin-structure factor, as a function of the magnitude of the jump operators. The variational results (red solid line) are compared to the MCWF method (blue dashed line).}
\label{fig4}
\end{figure}

We next consider variational states of the product-state type, $\rho = \rho_p = \prod_\veci \rho_{\veci}$ and minimize the upper bound of the norm
\eq
||\dot{\rho}|| \leq \sum_{\langle \veci \vecj\rangle} {\rm Tr}|\dot{\rho}_{\veci\vecj}| \to \min, \label{eq:min2}
\eqx
where the reduced two-site operators are defined as ${\dot{\rho}_{\veci\vecj} = {\rm Tr}_{\not{\veci}\not{\vecj}}{\,\dot{\rho}}}$. 
It is sufficient to minimize the norm $||\dot{\rho}_{\veci\vecj}||$ of a single bond, which for the case of an AF order can be expressed as
\eqn
||\dot{\rho}_{\veci\vecj}|| & \equiv & ||\dot{\rho}_{AB}|| = {\rm Tr}|\dot{\rho}_{AB}|, \\
\dot{\rho}_{AB} &=& -i [H_{AB}, \rho_{AB}] + \mathcal{D}_{AB}(\rho_{AB}) + \label{eq:norm} \\
&& \sum_{A'} {\rm Tr}_{A'}\left\{ -i [H_{BA'}, \rho_{ABA'}] + \mathcal{D}_{BA'}(\rho_{ABA'}) \right\}  + \nonumber \\
&& \sum_{B'} {\rm Tr}_{B'}\left\{ -i [H_{B'A}, \rho_{B'AB}] + \mathcal{D}_{B'A}(\rho_{B'AB}) \right\}.  \nonumber
\eqnx
Here A and B label the two sublattices and, e.g., $\rho_{AB} \equiv \rho_A \otimes \rho_B$, $\rho_{ABA'} \equiv \rho_A \otimes \rho_B \otimes \rho_{A'}$, while $\mathcal{D}_{AB}$ gives the dissipative part with the jump operators acting on the sites A and B. The first two terms of equation~(\ref{eq:norm}) correspond to an exact treatment of a single bond,  which already goes beyond the mean-field description, whereas the next ones describe interaction with the surrounding sites treated on the mean-field level, as visualized by the dashed lines in figure~\ref{fig4}(a).

In figure~\ref{fig4}(b) we compare the spin-structure factor obtained from our variational scheme and the MCWF method as a function of magnitude of the jump operators (which we set equal here without the loss of generality).
According to the results of both methods, the system exhibits substantial ordering (e.g. with spin-structure factor larger than $0.5$) when $\gamma_1 = \gamma_2 \gtrsim 1$. Although the variational method contains terms going beyond mean field, its results for AF ordering do not depend on the value of $U$, as opposed to the exact approach. This happens due to restriction of the density matrices to the form of product states. Moreover, the variational method overestimates the AF ordering due to the absence of fluctuations in our variational manifold. Therefore, the employed approaches are complementary to each other, and both indicate the formation of an AF order of substantial magnitude in the steady state.

\section{Conclusions}

In this paper, we proposed a scheme for dissipative preparation of antiferromagnetic order in ultracold fermions trapped in an optical lattice. We demonstrated that by using a combination of  two dissipative processes based on Raman-assisted hopping it is possible to engineer the dissipative dynamics in such a way that the AF phase emerges as  its dark state. By using a combination of an exact and variational approaches, we observed the formation of a strong AF order on the timescales achievable in present-day experiments.

We note that the technique presented here can be readily implemented in the setups already used to search for the AF order~\cite{Esslinger2015}, and thereby paves the way to an experimental realization of the AF phase in the Fermi-Hubbard model. While we exemplified the approach using the atomic level structure of $^{40}$K~\cite{Jordens2008},  the method is general and can  be also applied to other fermionic atoms currently available in laboratory, such as $^6$Li~\cite{Hart2015}, Er~\cite{AikawaPRL14}, Dy~\cite{LuPRL12}, Yb~\cite{PhysRevLett.105.190401}, and Cr~\cite{PhysRevA.91.011603}. After preparation of the AF phase with low entropy it should be possible to explore the phase diagram of the Hubbard model, including the pseudogap regime, by coherently removing a fraction of the atoms from the trap thereby introducing hole carriers into the system.
Finally, extending these ideas to single-site addressable lattices as offered by the fermionic quantum gas microscopes \cite{CheukPRL15, HallerNatPhys15, ParsonsPRL15, 2015arXiv151004744E}, opens the door to the preparation of more sophisticated many-particles states.

\acknowledgments

We acknowledge stimulating discussions with Ken Brown, Tommaso Calarco, Andrew Daley, Suzanne McEndoo, Tobias Osborne, Cindy Regal, Luis Santos, Micha\l{} Tomza, and Martin Zwierlein.
The work was supported by the People Programme (Marie Curie Actions) of the European Union's Seventh Framework Programme (FP7/2007-2013) under REA grant agreement n$^{\rm o}$ [291734], by the Volkswagen Foundation, and by DFG within SFB 1227 (DQ-mat).

\appendix

\section{Two-dimensional Jordan-Wigner transformation} \label{app:A}

The Jordan-Wigner transformation between fermionic operators $\hat{c}\dg_{\veci \s}$, $\hat{c}_{\veci \s}$ and spin operators ($\hat{\s}^{+}, \hat{\s}^{-}, \hat{\s}^{z}$) for spin-1/2 particles is defined as
\eqn
\hat{c}\dg_{\veci \s} &=& \hat{\s}_{k(\veci,\s)}^{+} e^{+i \pi \sum_{l<k(\veci, \s)} \hat{n}_l}, \label{eq:JW1} \\
\hat{c}_{\veci \s} &=& \hat{\s}_{k(\veci,\s)}^{-} e^{-i \pi \sum_{l<k(\veci, \s)} \hat{n}_l} \label{eq:JW2}, \\
\hat{c}\dg_{\veci \s} \hat{c}_{\veci \s} &=& \frac{1}{2} \left( \hat{\s}^z_{k(\veci, \s)} + 1 \right) \equiv \hat{n}_{k(\veci, \s)}, \label{eq:JW3}
\eqnx
where the function $k(\veci, \s)$ enumerates the fermionic particles for a given lattice site, $\veci = (i_x, i_y)$, and particle's spin, $\s = \ua, \da$. In this manner, fermions are positioned on a chain, where the position along the chain is given by $k(\veci, \s)$. The factor $e^{\pm i \pi \sum_{l<k(\veci, \s)} \hat{n}_l}$, where the summation runs over all fermions `before' the one at site $\veci$ with spin $\s$, evaluates to $+1 (-1)$ for even (odd) number of fermions `before' the given one. Thereby, the anticommutation rules for the fermionic operators $\hat{c}\dg_{\veci \s}$, $\hat{c}_{\veci \s}$ are fulfilled. We can also use the relation $e^{\pm i \pi \sum_{l<k(\veci, \s)} \hat{n}_l} = \Pi_{l<k(\veci, \s)} (-\hat{\sigma}^z_l)$, where $-\hat{\sigma}^z_l$ evaluates to $-1$ when the $l$-th spin is up (correspondingly, when there is a fermionic particle in the mode $l$) and to 1 in the opposite case.

\begin{figure}[t]
\begin{center}
\includegraphics[width=\columnwidth,angle=0]{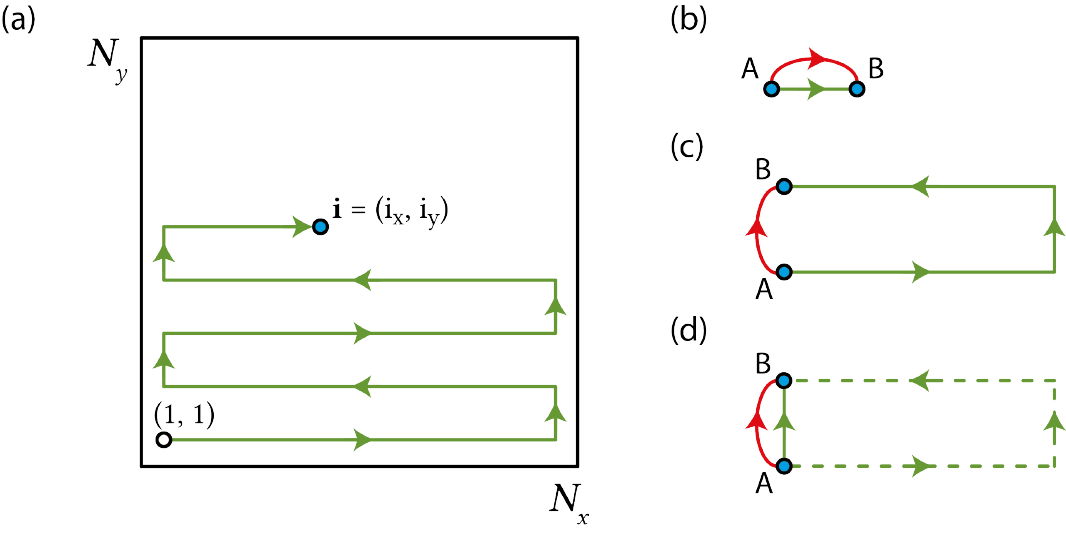}
\end{center}
\caption{Illustration of the Jordan-Wigner transformation for a $N_x \times N_y$ system. (a) Jordan-Wigner string (in green) from site $(1, 1)$ to site $\veci$. (b) and (c) illustration of the hopping terms and the related Jordan-Wigner strings along (b) and across (c) the 'rows' formed by the strings. (d) The hopping term as in (c) with approximation of the 'non-local' (dashed) part of the string. The remaining components of the string act only on the sites A and B (as visualized by the solid green arrow) and the hopping term has the same form as in (b).}
\label{figA1}
\end{figure}

For a one-dimensional system the function $k(\veci, \s)$ can be chosen simply as $k(\veci, \s) = 2 |\veci| + \delta_{\s,\da}$.
In two dimensions, however, there are a few possibilities to perform the Jordan-Wigner transformation~\cite{JWtrafo,Fradkin1989,PhysRevB.43.3786,PhysRevB.48.6136} (see Ref.~\cite{JWtrafo} for a review). Here, we consider an $N_x \times N_y$ system with lattice site coordinates $i_{x(y)} \in \{1, 2, ..., N_{x(y)}\}$ and use the following  function to enumerate particles (with~$m \in \mathbb{N}$)
\eq
k(\veci, \s) = \begin{cases}
      2\left[N_x (i_y-1) + i_x -1 \right] + \delta_{\s,\da}, & i_y = 2 m + 1, \\
      2\left[N_x (i_y-1) + N_x - i_x \right] + \delta_{\s,\da}, & i_y = 2 m.
   \end{cases}
\eqx
The chain of fermions and the related Jordan-Wigner string goes from left to right in the first row of the system, then in the second row goes to the left and forms a zig-zag (cf. figure \ref{figA1}(a)).

\section{Variational method for fermions} \label{app:B}

By performing the Jordan-Wigner transformation as described in appendix \ref{app:A}, we can  transform all terms of the master equation~(\ref{eq:master}), which we exemplify here by considering the hopping, $\hat{c}^\dagger_{\veci \s} \hat{c}_{\vecj \s}$. Using the transformation (\ref{eq:JW1})-(\ref{eq:JW3}) the hopping is expressed as
\eq
\hat{c}^\dagger_{\veci \s} \hat{c}_{\vecj \s} = \hat{\s}_{k(\veci,\s)}^{+} \left[ \Pi_{k(\veci, \s) < l < k(\vecj, \s)} (-\hat{\sigma}^z_{l}) \right] \hat{\s}_{k(\vecj,\s)}^{-}.
\eqx
If the hopping takes place between two sites in the same row, the Jordan-Wigner string is local (as illustrated in figure \ref{figA1}(b)). For example, \mbox{if $\vecj = \veci + (1, 0)$} and \mbox{$k(\veci, \s) < k(\vecj, \s)$}, we have
\eqn
\hat{c}^\dagger_{\veci \ua} \hat{c}_{\vecj \ua} &=& \hat{\s}_{k(\veci,\ua)}^{+} \left[-\hat{\s}^z_{k(\veci,\da)}\right] \hat{\s}_{k(\vecj,\ua)}^{-}, \label{eq:hop1} \\
\hat{c}^\dagger_{\veci \da} \hat{c}_{\vecj \da} &=& \hat{\s}_{k(\veci,\da)}^{+} \left[-\hat{\s}^z_{k(\vecj,\ua)}\right] \hat{\s}_{k(\vecj,\da)}^{-}. \label{eq:hop2}
\eqnx
If the hopping takes place between two sites in different rows, the Jordan-Wigner string includes also sites between the $\veci, \vecj$ pair and the (left or right) edge of the system (as illustrated in figure \ref{figA1}(c)). Note that the number of these `non-local' sites is always even due to our choice of the enumerating function $k(\veci, \s)$. Consequently, at half-filling, i.e. with one particle per site on average, there is an even number of particles, $2 N_1$, along the `non-local' part of the Jordan-Wigner string. In such case, the string evaluates to the factor $(-1)^{2 N_1} = 1$. In our variational method we use this value as an approximation for the `non-local' part of the string, whereas the local terms are retained. Such a procedure is similar to the `mean-field' treatment of the strings applied, e.g. in Refs.~\cite{JWtrafo,Derzhko2003407,PhysRevB.43.3786}. As a result, the hopping between rows is expressed in the same way as in equations (\ref{eq:hop1})-(\ref{eq:hop2}), which is visualized in figure~\ref{figA1}(d).

For the jump operators we use the same approximation. Thereby, the master equation~(\ref{eq:master}) acquires the form with at most two-site interaction terms (site here refers to the original fermionic site), for which the variational method can be readily applied. Note that in the MCWF calculations the `non-local' part of the strings needs to be preserved.

Explicitly, the contribution from the hopping term to the norm of a single bond, $||{\dot{\rho}_{AB}|| = ||{\rm Tr}_{\not{A}\not{B}}{\,\dot{\rho}}}||$, on the example of the first term in equation (\ref{eq:norm}), is given by
\begin{eqnarray}
-i [H_{AB}, \rho_{AB}] &=& -i t [\hat{\s}_{k(A,\ua)}^{+} \hat{\s}^z_{k(A,\da)} \hat{\s}_{k(B,\ua)}^{-}  \\
&& + \hat{\s}_{k(A,\da)}^{+} \hat{\s}^z_{k(B,\ua)} \hat{\s}_{k(B,\da)}^{-} + {\rm h.c.}, \rho_{AB}], \nonumber
\end{eqnarray}
where we assumed $k(A, \s) < k(B, \s)$. The treatment of other terms in the norm is analogous and, similarly, results in the appearance of parity factors ($-\hat{\s}^z_{k(A,\s)}$ and $-\hat{\s}^z_{k(B,\s)}$), which originate from the fermionic anticommutation rules.

Finally, let us note that the results of our variational method do not depend on the system size, $N_x \times N_y$ (however, they make sense only for $N_x, N_y \geq 4$) and hence correspond to the thermodynamic limit.

\bibliography{diss.bib}

\end{document}